\documentstyle[preprint,aps]{revtex}

\begin{document}
\draft
\preprint{}
\title{Spinodal Decomposition in Binary Gases}
\author{S. Bastea and J.L. Lebowitz}
\address{Department of Physics and Mathematics, Rutgers University,
Piscataway, New Jersey, 08855-0849}
\maketitle
\begin{abstract}
We carried out three-dimensional simulations, with about $1.4\times10^6$
particles, of phase segregation in a low density binary fluid
mixture, described mesoscopically by energy and momentum conserving 
Boltzmann-Vlasov equations. Using a combination of Direct Simulation
Monte Carlo(DSMC) for the short range collisions and a version of
Particle-In-Cell(PIC) evolution for the smooth long range interaction,
we found dynamical scaling after the ratio of
the interface thickness(whose shape is described approximately by a
hyperbolic tangent
profile) to the domain size is less than $\sim0.1$. The scaling length
$R(t)$ grows at late times like $t^\alpha$, with $\alpha=1$ for critical quenches and
$\alpha=\frac{1}{3}$ for off-critical ones. We also measured the
variation of temperature, total particle density and hydrodynamic
velocity during the segregation process.
\end{abstract}
\pacs{PACS numbers:  64.75.+g, 68.10.-m, 47.70.Nd}

\widetext
The process of phase segregation through which a system evolves towards
equilibrium following a temperature quench from a homogeneous phase
into a two phase region of its phase diagram has been of continuing interest
during the last decades \cite{gunton1},
but many problems still remain to be solved. 
This is particularly so for fluids,when particle,
momentum and energy densities are conserved locally; these are
currently the focus of both numerical studies 
\cite{ma1,valls1,frank1,wu1,bastea1,toxvaerd2} and  
micro-gravity experiments \cite{beysans1,perrot1}. 

In this paper we present computer simulations of spinodal
decomposition in a three-dimensional mixture of two kinds of particles that
we label 1 and 2 using a novel microscopic dynamics and computational
scheme. The particles interact with each
other through short range interactions modeled here by hard spheres having
the same mass $m$ and diameter $d$. Particles of different kinds
interact also through a long range repulsive Kac potential,
$V(r)=\gamma^{3}U(\gamma r)$. The equilibrium properties of such
a system are well understood, there is even a rigorous proof of a phase
transition at low temperatures to an immiscible state \cite{lieb},
which in the limit
$\gamma\rightarrow0$ \cite{lebowitz1}, is described by mean
field theory. 
When the density $n$
is low enough, $nd^3\ll1$, and the potential sufficiently long ranged,
$n\gamma^{-3}\gg1$, the free energy of the system is well approximated
by $F= 
k_BT\int[n_1(\vec r)\ln n_1(\vec r)+n_2(\vec r)\ln n_2(\vec r)]d\vec
r + \int V(|\vec r_1-\vec r_2|)n_1(\vec r_1)n_2(\vec
r_2)d\vec r_1 d\vec r_2$ and the $\gamma\rightarrow0$ critical
temperature, which should be an upper bound for $T_c^\gamma$ at
$\gamma>0$, is given by
$k_BT_c^0=\frac{1}{2}n\int U(r)d\vec r$. In this regime the 
dynamical evolution of the system should be well described by two 
coupled Boltzmann - Vlasov equations:
\begin{equation}
\frac{\partial f_i}{\partial t}+\vec v \cdot \frac{\partial
f_i}{\partial \vec r}+\frac{\vec F_i}{m}\cdot\frac{\partial
f_i}{\partial \vec v}=J[f_i,f_1 + f_2],\;\;\;\;\;i=1,2
\end{equation}
where $f_i(\vec r,\vec v,t)$ are the one-particle distribution functions, 
$\vec F_i(\vec r,t)=-\nabla\int V(|\vec r-\vec r\prime|)n_j(\vec
r\prime)d\vec r\prime$, $n_j(\vec r\prime)=\int f_j(\vec r\prime, \vec v,
t)d\vec v$ with $i,j=1,2$,$i\neq j$, and $J[f,g]$ is
the Boltzmann collision operator for hard core interactions \cite{chapman1}.
Kinetic equations of this type have been proposed in \cite{sobrino}, 
and if the system is quenched inside
the coexistence region they will describe gas-gas segregation
\cite{schouten} into two phases, one rich in particles
of type 1 and the other rich in particles of type 2. 
(Examples of gas mixtures that have a miscibility gap are
helium-hydrogen, helium-nitrogen, neon-xenon etc. \cite{schouten}.)
We believe that the model contains the essential features of phase 
separation in general binary fluid mixtures.

To simulate our system we modeled the Boltzmann collisional part using
a stochastic
algorithm due to Bird \cite{bird1}, known as Direct Simulation Monte
Carlo (DSMC), while for the Vlasov part we used the particle-to-grid-weighting
method, well known in plasma physics \cite{langdon1}.
In the DSMC method the physical space is divided into cells containing
typically tens of particles. The main ingredients of this procedure
are the alternation of free flow over a time interval $\Delta t$ 
and representative collisions among pairs of particles 
sharing the same cell. In the
particle-to-grid-weighting algorithm the particle densities are computed
on a spatial grid through some weighting depending on the particle
position, then the Vlasov forces are calculated on the same
grid. Finally, the forces at the position of each particle are
interpolated from the forces on the grid.
The coupling of these methods, which have been extensively 
used individually, made possible our simulations of phase segregation 
with $1.4\times 10^6$ particles, with only modest computational
resources: a typical run took about 32 CPU hours on a 233 MHz Alpha
Station.  It appears that this method can be extended to the study of
the effects of phase segregation on inhomogeneous hydrodynamical flows
of practical importance \cite{shell}.
 
Since one of our main interests was the late time hydrodynamical
regime, a delicate balance had to be struck between the size of the
system, the range of the potential, the temperature and the particle
density, making sure that each of the methods is used within its
range of validity and that their combination remains computer
manageable. On the one hand the potential must be reasonably long
ranged so that the Vlasov description is physically appropriate and 
numerically sound, and on the other hand it must have a range much smaller
than the size of the system. This restriction made necessary the use
of two spatial grids: a somewhat coarse one for the collisions and
a finer grid for the long range potential. It also imposed the use of
quadratic spline interpolation for the calculation of grid
quantities and a ten-point difference scheme for the calculation of the
forces \cite{langdon1}. 

Our results were obtained
using a system with 1382400 particles in a cube with
periodic boundary conditions. We also studied smaller
systems to identify unavoidable finite-size effects. The interaction 
potential used
was gaussian, $U(x) = \alpha \pi^{-\frac{3}{2}}e^{-x^2}$, $\alpha >
0$, but there is no reason to believe that different
repulsive potentials would qualitatively change the results.
All quenches were performed at a total particle density $nd^3\simeq0.01$
and an initial temperature $T_0$, $T_0/T_c^0=0.5$. 
The initial conditions for each run were random positions for all particles
and velocities distributed according to a maxwellian of constant
temperature.( In the DSMC evolution, as in the Boltzmann-Vlasov
equations, the hard cores only enter in determining the collision
cross sections.) 
The total energy of the system was very well conserved by the
dynamics. This meant that the kinetic energy and hence the temperature
increased as the system segregated, but at
late times it changed very slowly on the time scale of our
simulations. We indicate the final temperature T in the figures. The 
effective number of particles in the range of the potential was about 100-500.

In the following we compare results of our simulations
with available theoretical and experimental work 
and check various assumptions made in the former, 
e.g. the neglect of density and temperature variations.
We are also currently investigating both
formal and rigorous Chapman-Enskog and Hilbert expansion methods for
derivation of macroscopic evolution equations for this model \cite{bastea3}.
The units used for lengths and times are the mean-free path, 
$\lambda=(2^{\frac{1}{2}}\pi nd^2)^{-1}$, and mean-free time, 
$\tau=\lambda/c$, where $c=(2k_BT_0/m)^{\frac{1}{2}}$ and
$T_0=\frac{1}{2}T^0_c$ is the initial temperature.


We first present results for critical
quenches (equal volume fractions of the two species). Three different
potential ranges were used, and for each one of them 10-12
quenches were performed. 
The domain size was probed using the pair correlation function, 
$C(\vec r,t) = {\mathcal{V}}^{-1}<\sum_{\vec x}\phi(\vec x,t)\phi(\vec
x+\vec r,t)>$, 
where $\phi=(n_1-n_2)/(n_1+n_2)$ is the local order parameter, 
$n_1$ and $n_2$ are the local particle densities, $\mathcal{V}$ is
the volume and the average is over the different runs.
We determined $C(\vec r,t)$ by an inverse Fourier transform of the
structure function $S(\vec k,t)=|\tilde{\phi}(\vec k,t)|^2$; the order
parameter
$\phi(\vec r,t)$ and its Fourier transform $\tilde{\phi}(\vec k,t)$ were
computed on a $64\times64\times64$ cubic grid. The first zero of the 
spherically averaged correlation function, $C(r, t)$, was used as a
measure of the typical domain size, $R(t)$. The data are averages of 
the independent runs.

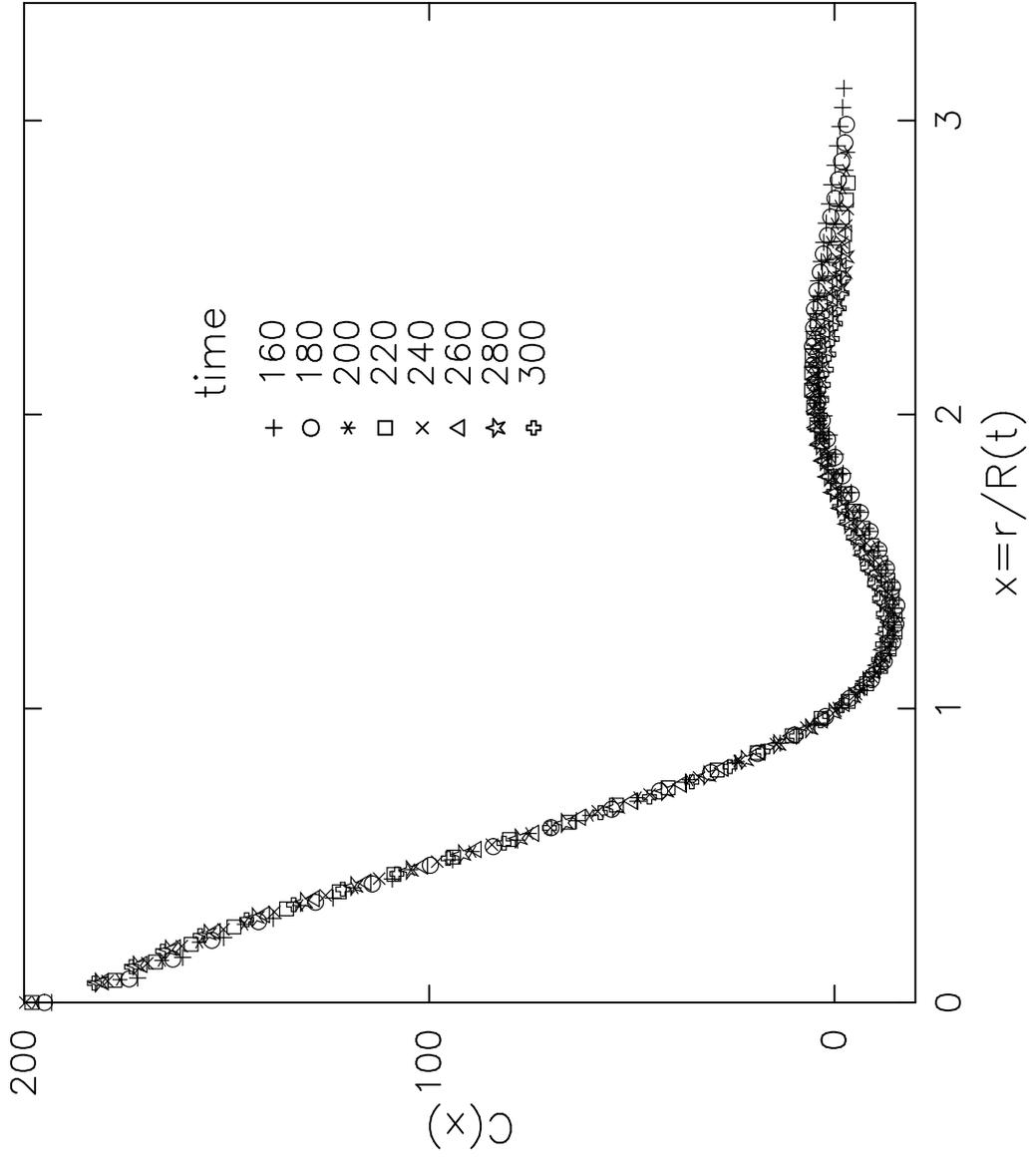
\begin{figure}
\caption{Scaled two-point correlation function for critical quench with
$\gamma^{-1}=0.4$. Final temperature $T=0.6T^0_c$.}
\label{corelation}
\end{figure}

Assuming the existence of a single characteristic length scale,
the dynamical scaling prediction \cite{gunton1} for the
late time spherically averaged correlation function is
$C(r,t)\simeq C(r/R(t))$. 
In Fig.\ \ref{corelation} the correlation function for 
potential range $\gamma^{-1}=0.4$ is plotted starting at
$t=160$, showing that the system is well within the scaling regime.

Simple dimensional analysis of the hydrodynamical evolution equations
in the limit of large domain sizes, appropriate for late times
, yields a linear growth law for the
domain sizes \cite{siggia1}, $R(t)\propto(\sigma/\eta)t$, when
$R$ is below $R_h=\eta^2/\rho\sigma$ and a $t^\frac{2}{3}$ law,
$R(t)\propto(\sigma/\rho)^\frac{1}{3}t^\frac{2}{3}$, for $R$
above $R_h$ \cite{furukawa2,bastea2}; $\sigma$ is the surface tension
coefficient, $\eta$ is the shear viscosity and $\rho$ is the density. 
We measured $\sigma$ directly using Laplace's law and
$\eta$ by studying the decay of a sinusoidal velocity profile.
The time evolution of $R(t)$ in our simulations is at late times (see
Fig. 2) $R(t)=a+b(\sigma/\eta)t$, with 
$b\simeq0.13\pm0.2$($\sigma=260$ and
$\eta=3250$ for $\gamma^{-1}=0.4$, $T/T_c^0=0.6$). This numerical factor
is similar to the one
observed in experiments \cite{beysans1} and recent large scale
molecular dynamics simulations \cite{toxvaerd2}. 
Whether or not a crossover to a $t^\frac{2}{3}$ growth occurs at
later times/larger domain sizes cannot be decided by our present results.
We estimated that we would need a system with at least 4 times
as many particles as the present one to be able to observe the
growth of domains with sizes bigger than $R_h$.

The linear regime starts around the time when
dynamical scaling begins to hold,
i.e. at a domain size of about 12-15 times $\gamma^{-1}$, the range of
the potential. 
This is in agreement with Siggia \cite{siggia1} who argued that the
linear regime is due to surface tension driven flows, so the
interfaces should be well defined, i.e. their width should be small
compared to the domain size. This width is usually taken to be of
order $\xi$, the correlation
length of order parameter fluctuations in the bulk phases. One expects
$\xi$ to be roughly equal to $\gamma^{-1}$, far away from the critical 
temperature, as we are\cite{rowlinson}. 

We also looked directly at the wall profiles separating different phases and 
found that they are approximately described by solitonic
solutions \cite{giacomin}. These are particle densities $n_i(z)$ depending on a
single spatial coordinate, such that
$f_i(z, \vec v)=n_i(z)exp(-mv^2/2k_BT)$ are stationary solutions of
Eq.(1) and $n_1(\pm\infty)=n_2(\mp\infty)$ are the mean-field equilibrium
densities at temperature $T$.
The order parameter profiles which satisfy the equation and the ones 
observed in the simulations(see Fig.\ \ref{wallp}) have both approximately the hyperbolic
tangent form \cite{rowlinson}, $tanh(z/2\xi)$, with $z$ the 
coordinate perpendicular to
the domain wall and $\xi$ a parameter which characterizes the
interface thickness. We found that this 
thickness is about $50$ percents bigger than $\gamma^{-1}$. The
total density is about $20$ percents smaller(for $\gamma^{-1}=0.4$,
$T/T_c^0=0.6$) at the interface
than in the bulk, in very good agreement with the solitonic solution
(see Fig.\ \ref{wallp}). The solitonic profiles were calculated at an 
effective temperature $T_{eff}$ for
which the asymptotic values of the order parameter matched the ones
observed in the simulations.  These asymptotically matched mean-field
profiles are steeper(in units of $\gamma^{-1}$) than the ones
observed, with better agreement as $\gamma$ is decreased. We believe
that this discrepancy is due to the fact that
$T/T_c^\gamma>T_{eff}/T_c^0$, as the equilibrium curve for $\gamma>0$ is
flatter around $T_c$ than the mean-field curve.

The need for a clear separation between boundary width and domain size
may explain why in earlier molecular dynamics 
simulations \cite{ma1}, a smaller 
exponent is found: in those computations the maximum domain size
observed is only 6-8 times the range of the potential, so the true
hydrodynamic regime was probably never reached.
In simulations using the lattice Boltzmann method
the interaction range is 
of the order of one lattice spacing and the 
hydrodynamic exponent is observed when the domain size is
about 10 lattice units \cite{frank1}, in agreement with our analysis.

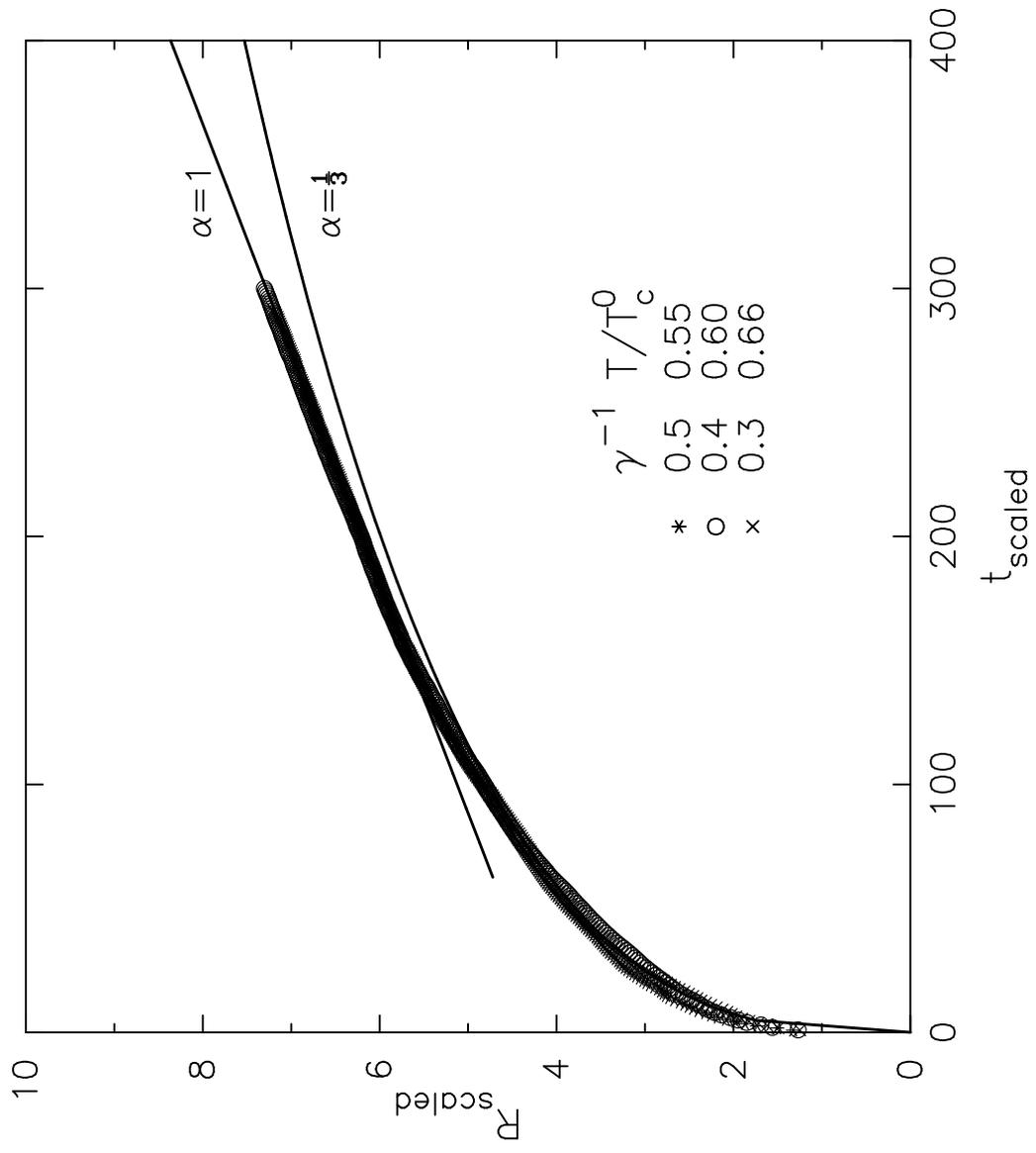
\begin{figure}
\caption{Scaled domain growth at critical quench; the $\gamma^{-1}=0.3$
and $\gamma^{-1}=0.5$ curves have been collapsed onto the
$\gamma^{-1}=0.4$ curve. A straight line fit($\alpha=1$) is drawn for
late times and a $\propto t^{\frac{1}{3}}$ fit($\alpha=\frac{1}{3}$) is drawn
for early times.}
\label{cscaling}
\end{figure}

At early times the growth is consistent with a $t^{\frac{1}{3}}$ behavior,
for all potential ranges\cite{siggia1,akcasu}. 
This exponent is not associated with
a scaling regime and it may not be universal, but it is not inconsistent
with the experimental results \cite{beysans1,perrot1}.
In fact, we were able to collapse the three curves onto
each other through scaling of the lengths and times Fig.\
\ref{cscaling} \cite{bastea1}.

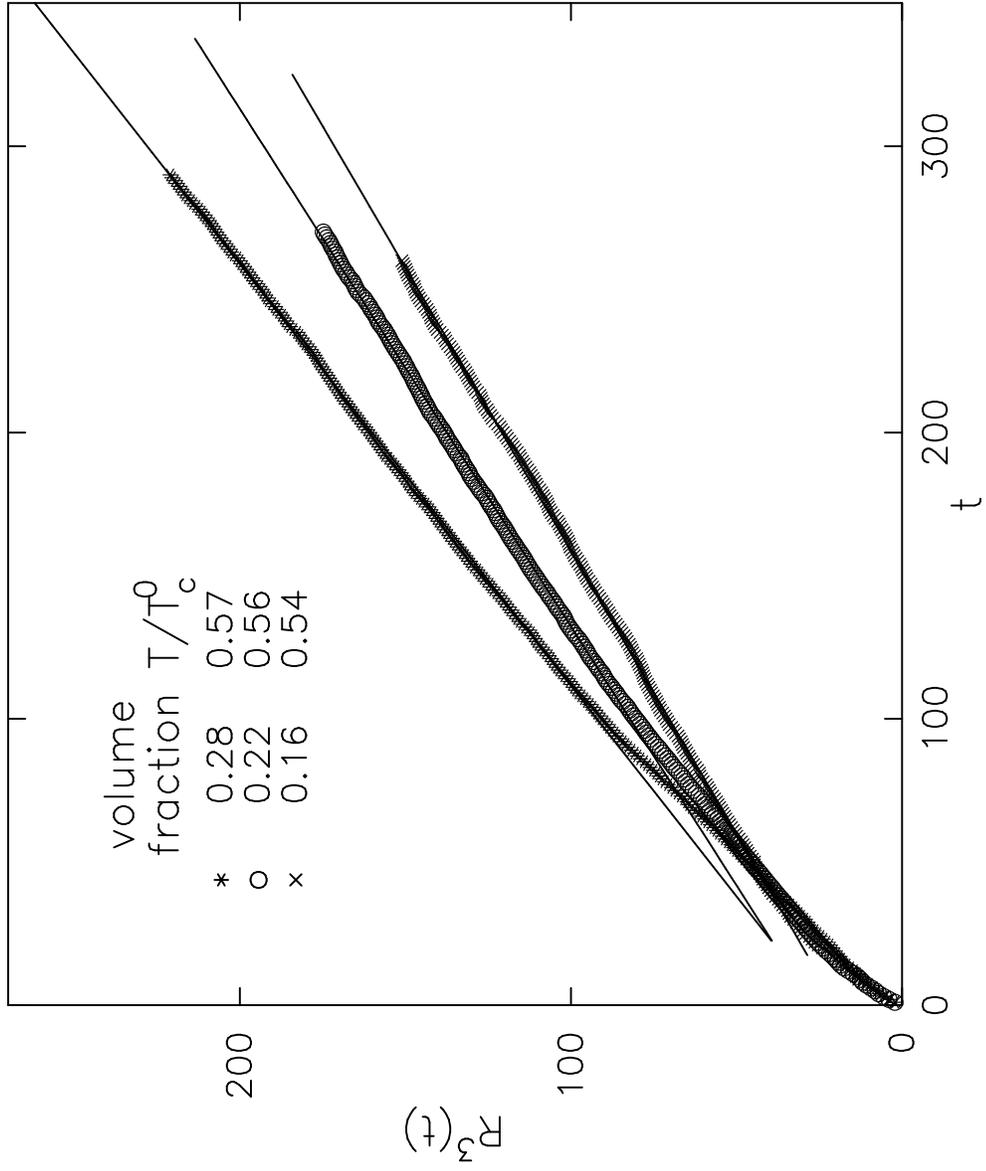
\begin{figure}
\caption{Domain growth for off-critical quenches with $\gamma^{-1}=0.4$;
straight line fits are drawn.}
\label{offc}
\end{figure}
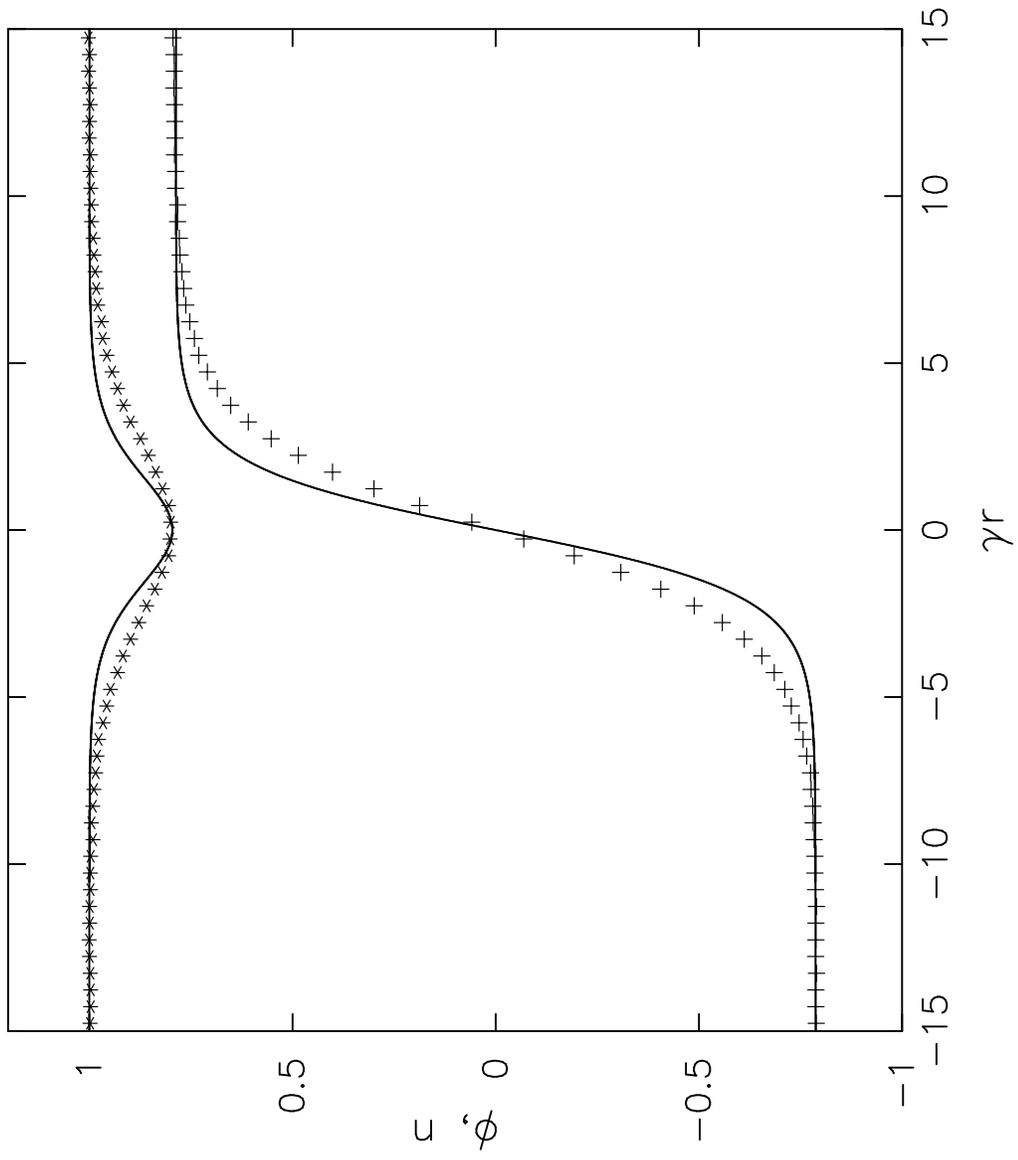
\begin{figure}
\caption{Total density $n(*)$ and order parameter $\phi(+)$
variations at the interface between the two phases;
$\gamma^{-1}=0.4$ and $T/T^0_c=0.6$. The full lines are the solitonic
solution(see text).}
\label{wallp}
\end{figure}

Off-critical quenches were performed for a single range of the
potential and three volume fractions. To our knowledge no
published results of such simulations exist, although they
are mentioned in \cite{toxvaerd2}. Therefore, we
compare our results to recent experimental work \cite{perrot1}
and analyze them using the known coarsening mechanisms
\cite{siggia1,vadim1}. In Fig.\ \ref{offc} we present the domain growth
for $\gamma^{-1}=0.4$ and volume fractions 0.16, 0.22 and 0.28. We plot
$R^3(t).vs.t$, and the late times regime is clearly
consistent with a $\frac{1}{3}$ exponent. In this regime the 
system satisfies dynamical scaling very well.
The presence of a late times $t^\frac{1}{3}$ growth at volume fractions
less or equal than about $0.3$ was observed recently in 
micro-gravity experiments
\cite{perrot1} and reasonably explained theoretically using a droplet
coalescence mechanism \cite{siggia1,vadim1}. This
regime was also analyzed by Siggia who predicted a prefactor
proportional to $v^\frac{1}{3}$($v$ being the volume fraction of the
minority species). In our simulations the prefactors are in reasonable
agreement with the above prediction. Furthermore, we can clearly see
the motion and coalescence of the droplets in movies of the dynamics.

In writing down evolution equations for the order parameter and
velocity fields in a symmetric binary system with momentum and energy
conservation it is generally assumed that the density and temperature
variations are small\cite{hohenberg1}.
We were able to check these
quantities directly by dividing the system into 
'hydrodynamical' cells, each containing about 100 particles, and computing
the order parameter, density, temperature($k_BT\equiv\frac{1}{3}m(\langle\vec
v^2\rangle-\langle\vec v\rangle^2)$, averages over the cell) and fluid velocity in each
cell. While the statistical fluctuations of these quantities are not 
vanishingly small, as they should be for true hydrodynamic variables, 
they are small enough for such a description to be at least
reasonable.
As already mentioned the total density is smaller at the boundary
between domains. Away from the domain boundaries the values of
the density and order parameter were close to their equilibrium
values. The temperature appeared to be uniform, with normal
equilibrium fluctuations everywhere in the system. 
This shows that the time scales over which heat
transport takes place are much smaller than the time scales
over which there is significant phase separation. 

If, as argued, the linear growth regime is due to surface tension
driven flows, one would expect bigger hydrodynamic velocities close to
the interfaces than in the bulk. 
We looked therefore at the distribution of hydrodynamic velocities as a
function of the order parameter for critical quenches.
As the system segregates, bigger hydrodynamic velocities
are typically observed close to the domain walls, identified by small
values of the order parameter. While this may be due in part to the
density being smaller at the boundary between domains,
it is also consistent with
the idea that at late times the flows are generated mainly by the
curvature of well formed interfaces. We are planning to study this issue
more carefully using bigger systems.

A natural refinement of our model would be the use of the Enskog
correction for the collisions \cite{chapman1}(which would require
numerically replacing the DSMC part with its recently proposed
extension \cite{frank2}, the Consistent Boltzmann Algorithm(CBA)) and
of long range attractive forces between like particles.

We believe that we have introduced a new model which is closer
to reality than lattice gases usually simulated, but still tractable
numerically by the new techniques that we have introduced. These
techniques are of independent interest and have enabled us to compute
temperature changes, density variations and profiles not done
before. Furthermore, we can compare our results to some exact ones. In
fact, we think that this model is a paradigm for the rigorous
derivation of hydrodynamic equations. 

We would like to thank Frank Alexander for useful discussions. This
research was supported in part by AFOSR Grant 0159 and 
NSF Grant 92-13424.

\end{document}